\documentclass[
  pre,
  twocolumn,
  twoside,
  byrevtex,
  superscriptaddress,
  floatfix,
  longbibliography,
]{revtex4-1}

\usepackage[realmainfile]{currfile}
\usepackage{color}

\definecolor{todoblue}{RGB}{0, 91, 187}

\usepackage{graphicx,epsfig,verbatim,enumerate}
\usepackage{amssymb,amsmath}
\usepackage{ifthen}

\usepackage{longtable}

\usepackage{mathtools}

\newboolean{twocolswitch}

\newcommand{\sindex}[1]{}
\newcommand{\nindex}[1]{}

\newcommand{\etal}{\textit{et al.}}
\newcommand{\www}[1]{\url{#1}}

\usepackage{lettrine}

\newcommand{\jsd}{D^{\textnormal{JS}}}

\setboolean{twocolswitch}{true}

\begin{document}

\title{\protect
Long-term word frequency dynamics derived from Twitter are corrupted:\\
A bespoke approach to detecting and removing pathologies in ensembles of time series
}

\author{
  \firstname{Peter Sheridan}
  \surname{Dodds}
}

\email{peter.dodds@uvm.edu}

\affiliation{
  Computational Story Lab,
  Vermont Complex Systems Center,
  MassMutual Center of Excellence for Complex Systems and Data Science,
  Vermont Advanced Computing Core,
  University of Vermont,
  Burlington, VT 05401.
  }

\affiliation{
  Department of Mathematics \& Statistics,
  University of Vermont,
  Burlington, VT 05401.
  }

\author{
  \firstname{Joshua R.}
  \surname{Minot}
}

\affiliation{
  Computational Story Lab,
  Vermont Complex Systems Center,
  MassMutual Center of Excellence for Complex Systems and Data Science,
  Vermont Advanced Computing Core,
  University of Vermont,
  Burlington, VT 05401.
  }

\author{
  \firstname{Michael V.}
  \surname{Arnold}
}

\affiliation{
  Computational Story Lab,
  Vermont Complex Systems Center,
  MassMutual Center of Excellence for Complex Systems and Data Science,
  Vermont Advanced Computing Core,
  University of Vermont,
  Burlington, VT 05401.
  }

\author{
  \firstname{Thayer}
  \surname{Alshaabi}
}

\affiliation{
  Computational Story Lab,
  Vermont Complex Systems Center,
  MassMutual Center of Excellence for Complex Systems and Data Science,
  Vermont Advanced Computing Core,
  University of Vermont,
  Burlington, VT 05401.
  }

\author{
  \firstname{Jane Lydia}
  \surname{Adams}
}

\affiliation{
  Computational Story Lab,
  Vermont Complex Systems Center,
  MassMutual Center of Excellence for Complex Systems and Data Science,
  Vermont Advanced Computing Core,
  University of Vermont,
  Burlington, VT 05401.
  }

\author{
  \firstname{David Rushing}
  \surname{Dewhurst}
}

\affiliation{
  Charles River Analytics,
  625 Mount Auburn Street,
  Cambridge, MA 02138.
}

\author{
  \firstname{Andrew J.}
  \surname{Reagan}
}

\affiliation{
  MassMutual Data Science,
  Amherst,
  MA 01002.
  }

\author{
  \firstname{Christopher M.}
  \surname{Danforth}
}

\affiliation{
  Computational Story Lab,
  Vermont Complex Systems Center,
  MassMutual Center of Excellence for Complex Systems and Data Science,
  Vermont Advanced Computing Core,
  University of Vermont,
  Burlington, VT 05401.
  }

\affiliation{
  Department of Mathematics \& Statistics,
  University of Vermont,
  Burlington, VT 05401.
  }

\date{\today}

\begin{abstract}
  \protect
  Maintaining the integrity of long-term data collection
is an essential scientific practice.
As a field evolves,
so too will that
field's measurement instruments and data storage systems,
as they are invented, improved upon, and made obsolete.
For data streams generated by opaque sociotechnical systems which
may have episodic and unknown internal rule changes,
detecting and accounting for shifts in historical datasets
requires vigilance and creative analysis.
Here, we show that around 10\% of day-scale word usage frequency time series for Twitter
collected in real time for a set of roughly 10,000 frequently used words
for over 10 years
come from tweets with, in effect, corrupted language labels.
We describe how we uncovered problematic signals while comparing
word usage over varying time frames.
We locate time points where Twitter switched on or off
different kinds of language identification algorithms,
and where data formats may have changed.
We then show how we create a statistic for identifying and removing words with
pathological time series.
While our resulting process for removing `bad' time series
from ensembles of time series is
particular, the approach leading to its construction
may be generalizeable.

\end{abstract}


\maketitle

\section{Introduction}
\label{sec:introduction}

The successful collection, cleaning, and storage of data through time requires
a stability of data sources, measurement instruments, and data storage taxonomy~\cite{osgood1957a,thurstone1959a,gooday1990a,gould1996a,mohr2008a,golinski2008a,pfeffer2018a,flake2019a}.
Of course, such stability has hardly been the norm for any developing area
of measurement.
Indeed, over the full arc of science,
measuring and recording time itself.
Thousands of years led to the establishment of a settled calendar,
with its quadracentennial
leap-year exception
to an exception
to an exception~\cite{eco1989a,gould1999a}.
Accurate clocks only first appeared
with chronometers in the 1600s~\cite{sobel2007a},
now, in terms of achievement, perhaps manifested by
the Global Positioning System (GPS)
which requires general relativity.

For internet data, sources go through episodic upgrades as 
formats are reconfigured and expanded.
In the case of Twitter, our focus here,
just a few of the features that have been added include:
retweets as formalized entities,
images and video,
local time,
and
tweet and user language.
The data object behind any given tweet,
whose format began as xml and changed to json,
has correspondingly grown in size, and the format
has evolved somewhat biologically.
The json for a ``quote tweet'' contains
simplified json for the retweeted tweet.
And the expansion from 140 to 280 characters
was accomplished not by expanding an existing
entry field but adding a second one which
must be combined with the old one for ``long tweets''.
Data providers and APIs have also have also changed, most recently to
GNIP as the data provider with a completely different JSON schema.

Over time, and not without setbacks,
Twitter has become an important global social media service.
Amplifying and reflecting real world stories,
Twitter is globally entrained with
politics and news,
sports,
music,
and
culture,
and also performs as a distributed sensor system for natural disasters
and emergencies~\cite{murthy2018a,
  sakaki2010a,
  lampos2010tracking,
  Culotta2010towards,
  hong2011a,
  younus2011average,
  bollen2011b,
  pickard2011time,
  gao2011harnessing,
  kursuncu2019a,
  grinberg2019a,
  gallagher2019a,
  nakov2019a,
  jackson2020a,
  tangherlini2020a,
  allen2020a,
  steinert2015online}.
Like any scientific enterprise, empirical research involving Twitter
and social media in general depends fundamentally on the quality of data~\cite{pfeffer2018a}.
Because of Twitter's now sprawling platform across time and language,
great care must be taken to ensure such integrity.

In this short report, we describe:
(1) How we uncovered
anomalies in word usage time series derived from Twitter
(Sec.~\ref{sec:storyturb.forensics}),
and
(2) One approach to identifying and removing corrupted time series
(Sec.~\ref{sec:storyturb.cleaning}).
We offer concluding thoughts in
Sec.~\ref{sec:storyturb.concludingremarks}.

We emphasize that we are not attempting to clean individual time series,
a common statistical practice, but rather
we are cleaning ensembles of times series by removing problematic,
unsalvagable time series.
Our work would be suitable for any many-component complex system
where abundances of components are recorded over time.

\begin{figure*}[t]
  \centering
  \includegraphics[width=0.90\textwidth]{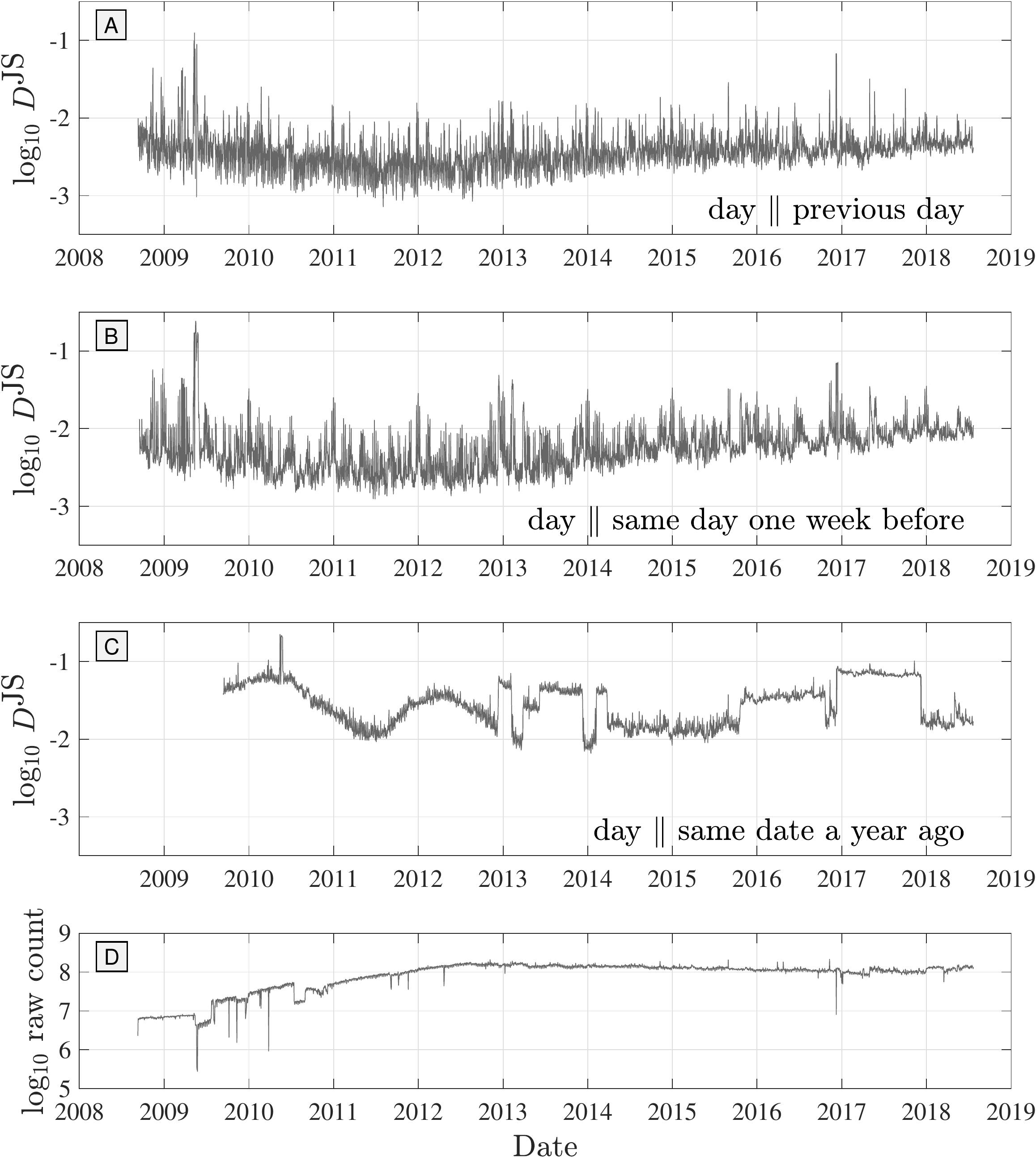}
  \caption{
    \textbf{A--C.}
    Three Jensen-Shannon (JS) divergence ($\jsd$) time series
    for day-day comparisons of word usage frequency distributions
    for the labMT word list~\cite{dodds2011e}
    over times scales of 1 day, 1 week, and 1 year.
    In the second panel (B), for example, each point gives the
    $\jsd$ for the labMT word frequency distributions of a day
    and for the same day a week earlier.
    Increases in story turbulence are suggested by
    the $\jsd$ time series for the day and week comparisons
    (\textbf{A} and \textbf{B})
    which both show slow increases from around 2011/2012 on.
    The day-same-date-a-year-before $jsd$ time series (\textbf{C})
    has peculiar jumps indicating the underlying time series of labMT
    words are corrupted in some way.
    \textbf{D.}
    Raw counts in the labMT data set as a function of date
    derived from an approximate 10\% feed of tweets from Twitter.
    There are clear jumps and drops in volume and these reflect changes in the feed rather
    than the collection process.
    None of the the dates of sharp transitions correspond with those presented
    by Jensen-Shannon divergence ($\jsd$) for year-scale comparisons in
    panel \textbf{C}.
    One explainable spike that does occur in
    \textbf{A}, \textbf{B}, and \textbf{C}
    is due to Twitter's entire systems failing for around a week in May of 2019.
    The jumps in $\jsd$ turn out to due to changes
    in Twitter's language detection algorithm.
  }
  \label{fig:storyturb.storyturbulence_twitter_labMT_jsd_daily001}
\end{figure*}

\begin{figure*}[t]
  \centering	
    \includegraphics[width=\textwidth]{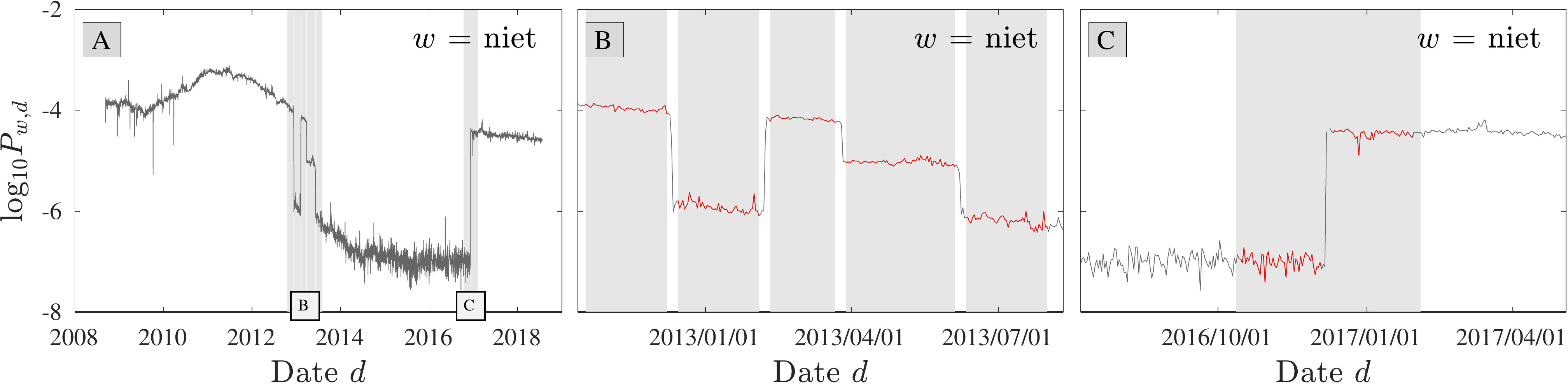}
    \caption{
      \textbf{A.}
      Normalized usage frequency time series for the Dutch word `niet',
      an example of a word strongly showing the effect of Twitter
      episodically altering their language detection algorithm.
      Dutch words were especially susceptible to being
      misclassified as being English giving rise to a corrupted time series.
      We expand the time series for `niet'
      in the two regions shaded in gray
      in panel \textbf{A}
      and present them in panels
      \textbf{B}
      and
      \textbf{C}.
      The jumps in the time series in panel \textbf{B}
      appear to be due to Twitter putting into place
      a series of language identification algorithms (which we do not attempt to reverse engineer in any way).
      The second jump in panel \textbf{B}
      seems to be due to the initial algorithm being switched off.
      The time series for `niet' stays roughly two orders of magnitude lower
      for over three years before one last major adjustment
      in late 2016 shown in panel \textbf{C}.
    }
    \label{fig:storyturbulence_twitter_labMT_show_count_jumps001}
\end{figure*}

Our approach is intended to be used for ensembles of time series
which can only be taken as they are, i.e., they cannot be rebuilt
from more primary data sets.
The work we present here has inspired 
a ground-up, re-identification of language for our Twitter data set~\cite{alshaabi2020a}
which in turn has led to the building of our $n$-gram time series for Twitter project
Storywrangler~\cite{alshaabi2020d,storywrangler2020-08-23a},
and the
revision and expansion
of our Hedonometer instrument~\cite{dodds2011e,dodds2015a,hedonometer2020-08-23a} (more below).
Our work is also
connected to our studies of
how the COVID-19 pandemic has been discussed across languages on Twitter~\cite{alshaabi2020b,dewhurst2020a},
as well as story turbulence and chronopathy in connection with Trump~\cite{dodds2020n}.

Throughout, we do not attempt to reverse engineer any
of Twitter's proprietary algorithms, but rather contend
with derived data and changing formats only.
Neither do we suggest that there is any fault of Twitter in changing
language identification methods, or indeed any aspect of their service, over time.
We also acknowledge that some data artifacts may have been
introduced by our own struggle with the
complexities of consistently processing formats that have changed many
times.

\section{Uncovering the presence of corrupted time series}
\label{sec:storyturb.forensics}

The instigation of our work here came from first noticing in June of 2018 that our
Hedonometer's happiness time series for English Twitter~\cite{dodds2011e,dodds2015a,hedonometer2020-08-23a}
had begun to apparently show increasing turbulence from the year 2016 on.
While a weekly cycle had always been a feature of our measure of Twitter's day-scale happiness
(Saturday had been typically happiest, Tuesday the least), its strength
appeared to be waning.

Deciding that this observation deserved further investigation,
we began to conceive of ways to measure
lexical and story turbulence~\cite{pechenick2017a,dodds2020n}.

Our Hedonometer functions by averaging
the individual, offline-crowd-sourced happiness scores.
At that point in time, we were using
a ``lexical lens'' of 10,222 words to create a single score for each day~\cite{dodds2011e}.
In brief, our method ultimately derived from Osgood \etal's
work on the measurement of meaning~\cite{osgood1957a}.
Through semantic differentials, Osgood \etal\ found
that valence (happiness-sadness) was the first dimension
of the experiences of meaning, followed by excitement and dominance.
Using a double-Likert scale, we improved upon earlier efforts
to score individual words~\cite{bradley1999a}, drawing on
the most common words used for various time periods
of Twitter, Google Books, the New York Times, and music lyrics~\cite{dodds2011e}.
We scored 10,222 words using Amazon's Mechanical Turk crowd-sourcing service,
calling the resulting data set labMT (language assessment by Mechanical Turk).
To run the Hedonometer, we created a usage frequency distribution for
this set of 10,222 labMT words, 
doing so for each day (according to Coordinated Universal Time) using tweets identified
as English by Twitter.

For an initial attempt to quantify turbulence on Twitter, we left
the Hedonometer part aside and focused on the underlying labMT word frequency distributions.
We used Jensen-Shannon divergence (JSD) to compare frequency distributions between dates
over different time scales,
with the distributions normalized as probabilities (or rates).
Our choice of Jensen-Shannon divergence was not crucial,
but rather something to try,
and we later developed alternate kinds of divergences
(see Refs.~\cite{dodds2020a,dodds2020g}).

In Fig.~\ref{fig:storyturb.storyturbulence_twitter_labMT_jsd_daily001}A--C,
we show three JSD time series representing comparisons between
a date and
A.\ the previous day,
B.\ the same day of the week, a week before,
C.\ and the same date, a year before.
We first plotted just the panels in 
Fig.~\ref{fig:storyturb.storyturbulence_twitter_labMT_jsd_daily001}A
and
Fig.~\ref{fig:storyturb.storyturbulence_twitter_labMT_jsd_daily001}B,
and saw that these JSD time series, after trending down from 2009 through to 2011,
were both increasing from 2012 on,
in agreement with our visual observations of Hedonometer.

In seeking to further develop our analysis of lexical turbulence, we
then examined JSD over longer time scales between dates,
including the year scale of
Fig.~\ref{fig:storyturb.storyturbulence_twitter_labMT_jsd_daily001}C.
And it was here that we first clearly saw there were problems with our
word distributions.
In late 2012, through 2013, and into 2014, we see striking jumps in
year-scale JSD.
We see more isolated jumps at the ends of 2015, 2016, and 2017.
Because we are comparing across years, we expect the anomalous
patterns to appear twice with year separation,
once for a problematic date looking back a year, and then
again for a year ahead looking back at the same problematic date.

We were able to say something immediately about what these anomalies are not.
They are not due to isolated corrupted dates, something we would have to contend
with in collecting any form of streaming data, as we would see these as spikes in the JSD.
Some aspect of the distributions was being switched and maintained.
Nor are the changes somehow volume dependent,
as Fig.~\ref{fig:storyturb.storyturbulence_twitter_labMT_jsd_daily001}D
makes clear.
While we do have some inconsistencies and changes in the volume of labMT words
collected over time, they do not line up with the jumps in the year-scale JSD time series.
While Twitter is ever-changing in content,
we nevertheless expect to find reasonable
consistencies in aggregate word usage patterns we may derive.

Upon visual inspection of individual
frequency time series for Twitter around the dates of the jumps
in the year-scale JSD time series, we find some
corresponding peculiar jump sequences.
(In the following section, we develop a systematic approach
to identifying such anomalous time series.)

For an individual example,
in Fig.~\ref{fig:storyturbulence_twitter_labMT_show_count_jumps001}A,
we show how the normalized usage frequency for the Dutch word `niet'
(English: `not/no') exhibits a number of sharp jumps (shaded regions).
The
word usage rate for `niet' increases or drops over several orders magnitude
around certain dates.
Expanding the shaded regions of
Fig.~\ref{fig:storyturbulence_twitter_labMT_show_count_jumps001}A,
Fig.~\ref{fig:storyturbulence_twitter_labMT_show_count_jumps001}B
shows four jumps occurring at the end of 2012 and in 2013, and
Fig.~\ref{fig:storyturbulence_twitter_labMT_show_count_jumps001}C
shows one in late 2016.

We have the suggestion then that individual tweets (and hence words)
are being differentially classified
by a sequence of language identification
algorithms employed by Twitter.
Overall, from Fig.~\ref{fig:storyturbulence_twitter_labMT_show_count_jumps001},
the example word of `niet'
seems to be initially identified as coming from English tweets, then,
after a several months of algorithms switching on or off,
appears to have been excluded from English for several years until
the end of 2016, or appear so due to a change to the tweet distribution
system provided by GNIP.

For the Hedonometer, for which these time series were prepared, we
had accepted tweets for processing unless they were identified as being a language
other than English or the user a speaker of language different from English
(in other words, not not English tweets).
We note that we had not noticed any of the year-scale JSD artifacts in our Hedonometer signal,
which itself is a day-scale average.

Word usage distributions are of course determined within the context of all
words for each day.
Given the behavior of year-scale JSD
in Fig.~\ref{fig:storyturb.storyturbulence_twitter_labMT_jsd_daily001}C,
we must expect the time series
of more words to follow the specific form of
`niet'.
We should also expect that these corrupted time series would
also lead to corrupted time series of basic function words in English (e.g., `the').

\section{Bespoke detection and removal of corrupted time series}
\label{sec:storyturb.cleaning}

\begin{figure}[tp!]
  \centering	
    \includegraphics[width=0.95\columnwidth]{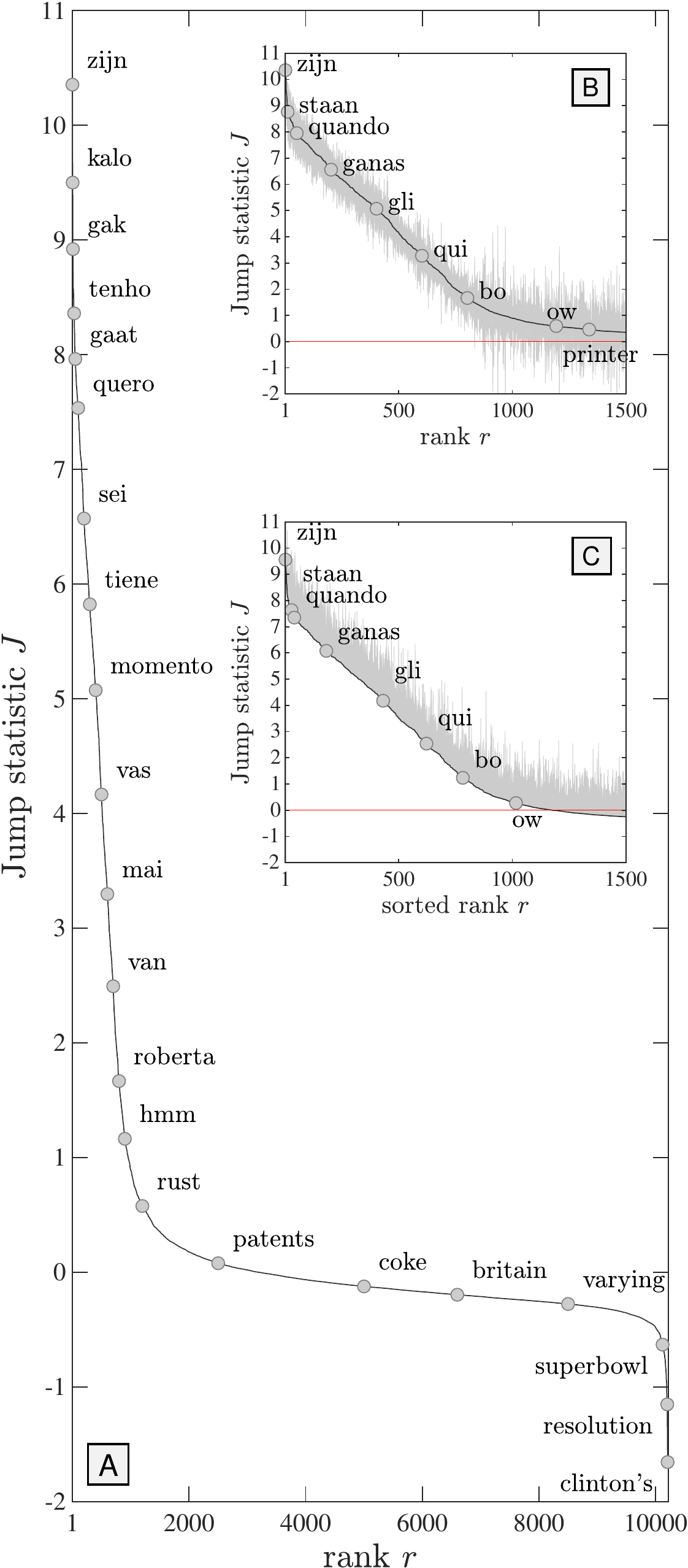}
    \caption{
      \textbf{A.}
      Sorted jump statistic $J$ with example words annotated.
      Words with high values of $J > 0$ are candidates for corrupted time series
      to be removed from the overall ensemble.
      \textbf{B.}
      The same plot as \textbf{A} but now for the 1,500 words
      with the highest value of $J$ and with the 95\% confidence
      interval $[J-2\sigma,J+2\sigma]$ marked in gray.
      \textbf{C.}
      The same plot as \textbf{B} with words
      re-ordered by descending values of $J - 2\sigma$.
      We take $J - 2\sigma > 0$ to be a criterion for
      a word's time series to be corrupted.
                      }
    \label{fig:storyturbulence_twitter_labMT_jump004_words}
\end{figure}

Clearly we do not want to involve poorly sampled time series in any of
our analyses.  And because we have observed that some words
follow the `niet' pattern while the majority track well (i.e., largely
continuously if noisy, and with jumps that have historical
explanations), we can hope to remove this particular set of poorly
sampled words.
We are thus able to overcome Twitter's hidden
shifts in algorithmic classification, at least in this most essential task of
extracting basic word usage frequency time series.

We construct a specialized method for
identifying corrupted time series as follows.
For the five jumps overall for `niet'
in Fig.~\ref{fig:storyturbulence_twitter_labMT_show_count_jumps001},
we notice that adjacent and intersticial time periods
are relatively quiescent.
Observing that similar patterns hold for other words,
we construct a ``jump statistic'' to measure the degree
to which a word's time series locally tracks the shapes in
Fig.~\ref{fig:storyturbulence_twitter_labMT_show_count_jumps001}B
and
Fig.~\ref{fig:storyturbulence_twitter_labMT_show_count_jumps001}C.

For the four jumps in the first time period of change
(Fig.~\ref{fig:storyturbulence_twitter_labMT_show_count_jumps001}B),
we choose five similar-length time ranges within which we expect words to be relatively
similar in abundance on a logarithmic scale:
\begin{center}
  \begin{tabular}{c}
    2012-10-19 to 2012-12-08 (51 days), \\
    2012-12-15 to 2013-02-03 (51 days),\\
    2013-02-10 to 2013-03-22 (41 days), \\
    2013-03-29 to 2013-06-04 (68 days), \\
    and \\
    2013-06-11 to 2013-07-31 (51 days). \\
  \end{tabular}
\end{center}
Again referring to the behavior of `niet',
we expect the transitions of corrupted words between these time periods to
be down, up, down, and down.

For the second time period (Fig.~\ref{fig:storyturbulence_twitter_labMT_show_count_jumps001}C),
we bound the one jump with two periods:
\begin{center}
  \begin{tabular}{c}
    2016-10-15 to 2016-12-04 (51 days), \\
    and \\
    2016-12-11 to 2017-01-30 (51 days).\\
  \end{tabular}
\end{center}
We expect corrupted words to jump up across this single transition.

For each word $w$ in our set of 10,222 words,
we construct a jump statistic $J$ by averaging
across differences of
the logarithms of normalized frequency
$P_{w,d}$ for all possible pairs of dates across each transition point.
We incorporate the expected transition direction for corrupted
time series by multiplying by +1 (up) or -1 (down), as appropriate.
By using sums of differences of logarithms,
we are equivalently computing ratios of normalized frequencies
and taking their geometric mean.

\begin{figure}
  \includegraphics[width=\columnwidth]{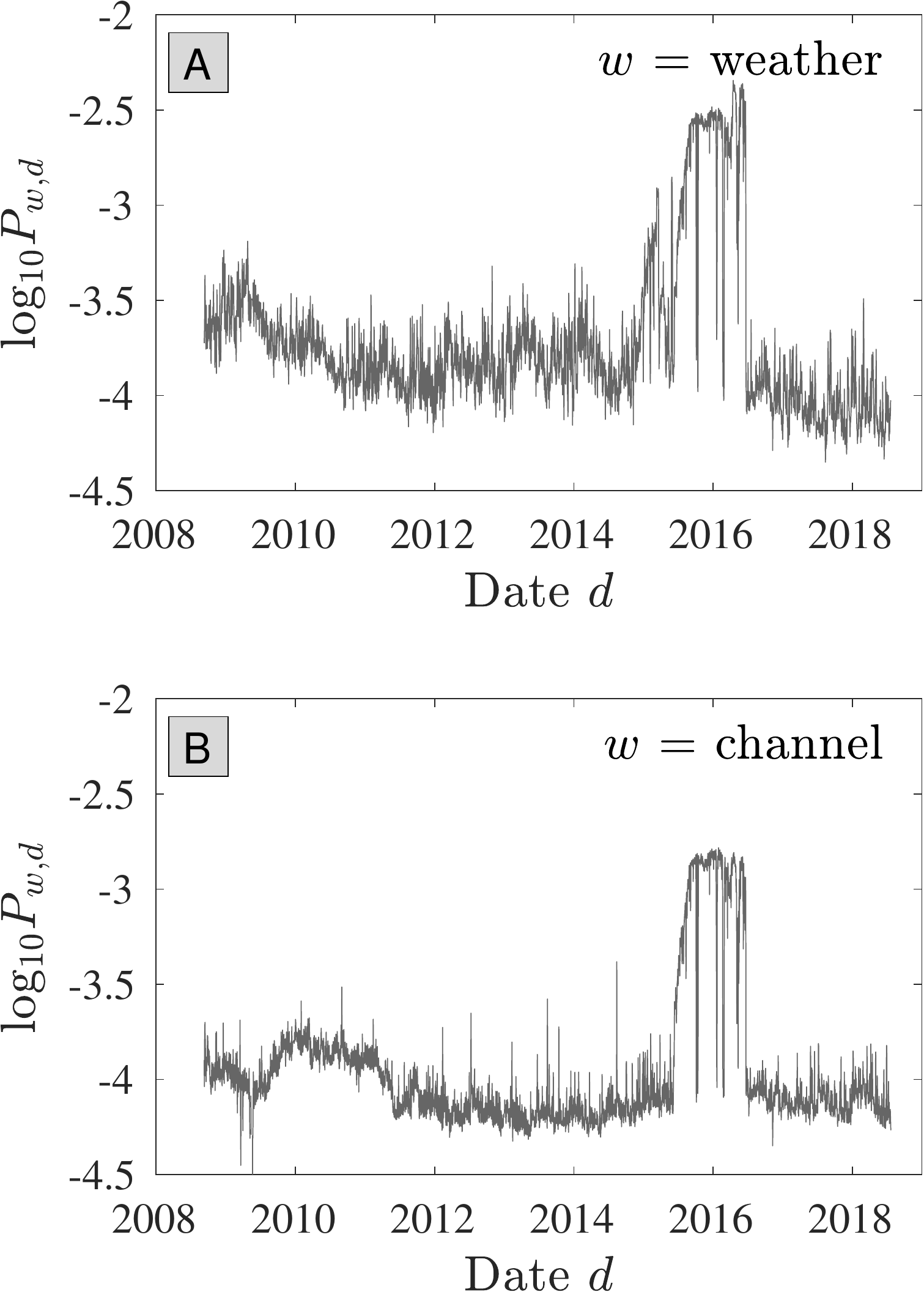}
  \caption{
    Two evidently connected words, ``weather'' and ``channel'',
    maintained an anomalous elevation in usage frequency for around a year, ending in
    November 2016.
    These words proved to induce an anomaly in the JSD time series for
    dates separated by one year, warranting their removal.
  }
  \label{fig:storyturb.storyturbulence_twitter_labMT_counts010}
\end{figure}

\begin{figure*}[tp!]
  \centering	
    \includegraphics[width=0.90\textwidth]{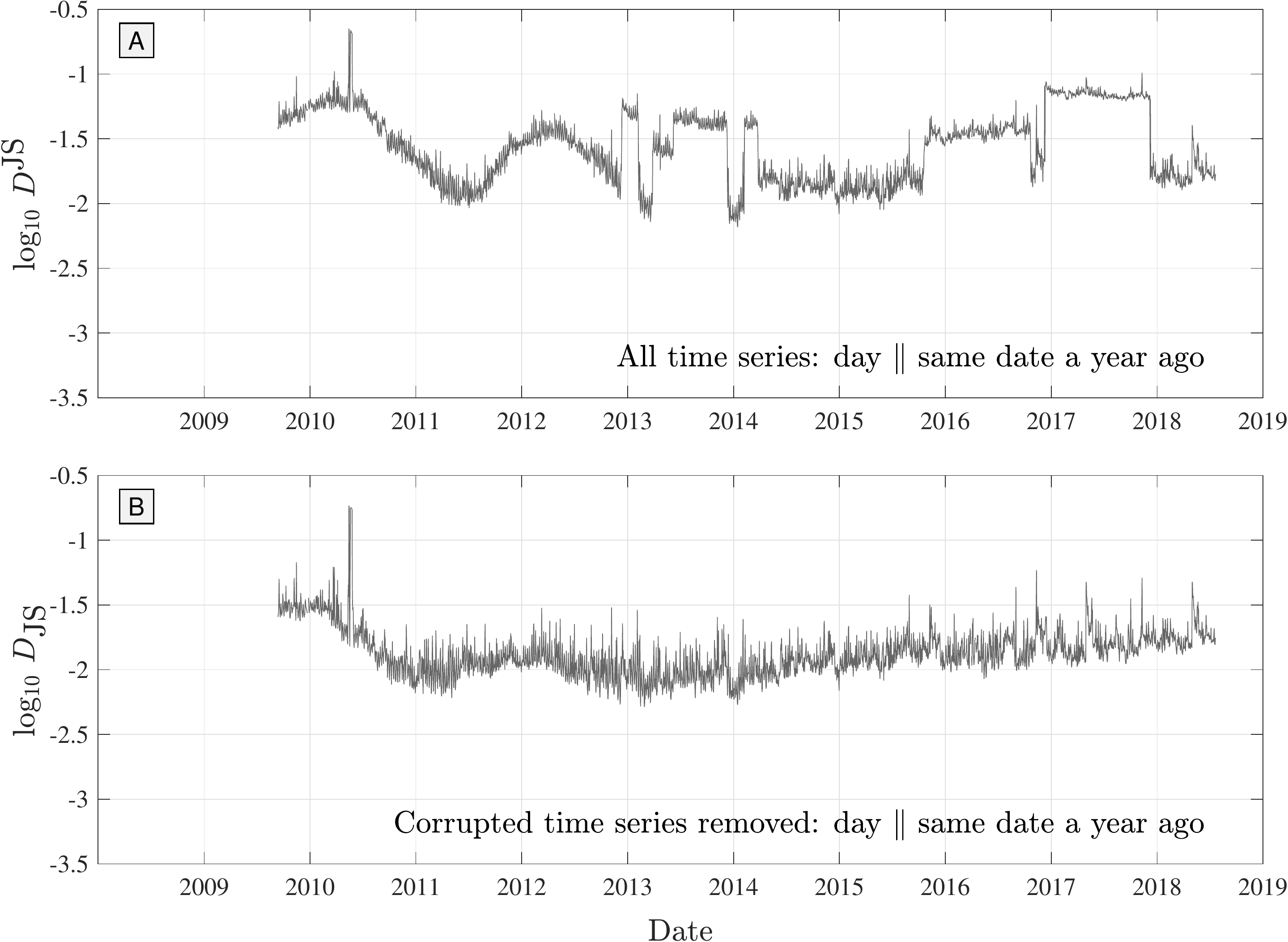}  
    \caption{
      \textbf{A.}
      For easy of comparison, a repeat of the year-scale JSD for labMT words
      presented panel \textbf{C}
      of Fig.~\ref{fig:storyturb.storyturbulence_twitter_labMT_jsd_daily001}.
      \textbf{B.}
      The same year-scale JSD times series but now performed on the labMT Zipf distributions
      with corrupted words removed.
  }
  \label{fig:storyturb.storyturbulence_twitter_labMT_jsd_daily007}
\end{figure*}

A simpler estimate might be to take the average probability of a word
in each region and sum the signed differences across the transition points.
However, comparing each pair of dates around each transition point
generates a distribution of $J$ values, allowing us to estimate
other statistics, such as a variance.

We compute a variance for each word by creating a distribution of values.
For each component of $J$ around each transition point (one value for each pair of dates).
For example, the first two time periods of 51 days each give us 2601 possible date pairs.
We use these to estimate variances for individual jumps.
We then sum variances over all five transition points
to obtain a variance for $J$ which we will denote simply by $\sigma^{2}$.

We compute $J$ and $\sigma^{2}$ for each word $w$.
We first sort them by descending values of $J$, and the main
plot in
Fig.~\ref{fig:storyturbulence_twitter_labMT_jump004_words}A
shows these values of $J$ for all 10,222 labMT words.
Annotated disks along the curve give example words.
We see that for positive values of $J$, the words that track with
the corrupted form are non-English words (`zijn', `kalo', `gak', etc.) and come
from a range of languages.
We also find corrupted time series for common words that tend to
be used across languages, such as ``hahaha''.

Visually, it appears that many of the
words ($\sim$ 90\%) have values of $J$ close to 0 (between -1/2 and 1/2, say).
These are non-corrupted words (`coke', `britain', and `varying').
We will firm up our measure of closeness below.

Some words go strongly against the trend of word corruption ($J < -1$)
with `clinton' and `hillary'
being prominent examples.
Twitter changed their language identification
algorithm about a month after the US presidential election,
and Clinton's loss led to her name dropping in prevalence,
going against the grain of jumping upwards for corrupted word time series
(Fig.~\ref{fig:storyturbulence_twitter_labMT_show_count_jumps001}C).

Now, having $J>0$ is too severe a condition for determining
whether a word is corrupt or not.
In the insets of
Fig.~\ref{fig:storyturbulence_twitter_labMT_jump004_words}B
and
Fig.~\ref{fig:storyturbulence_twitter_labMT_jump004_words}C,
we employ our distributions of $J$ scores to craft a better
criterion.

In Fig.~\ref{fig:storyturbulence_twitter_labMT_jump004_words}B,
we show the first 1500 words ordered again by decreasing $J$
but now with the range $J-2\sigma$ to $J+2\sigma$ shaded.
We observe words with $0 < J < 1$ whose
(notional) 95\% confidence interval covers 0.

Evidently, we would not want to exclude these words, mistaking them
for being corrupted because $J > 0$.
We will instead take our criterion for a time series to be corrupted
if $J-2\sigma > 0$.
In Fig.~\ref{fig:storyturbulence_twitter_labMT_jump004_words}C,
we re-order words so that they are descending according
to the lower limit of their 95\% confidence interval, $J-2\sigma$.
We preserve the example labeled words from
Fig.~\ref{fig:storyturbulence_twitter_labMT_jump004_words}B
to show how they move around.

With this criterion, we find that the time series
of 9,030 of our 10,222 word are relatively unaffected by
the five major changes in Twitter's language detection algorithm
we have identified.
We deem 1,192 words to be sufficiently problematic that
we should exclude them.

With these words removed, we return to our JSD calculations and
examine how the year-scale JSD now behaves.
We find that the jumps that appeared to be due to Twitter's language
detection algorithm changes have all been eliminated.

However, one last peculiar structure remained due to anomalous
word frequency changes in 2015 and 2016.
We were able to find that two words, ``weather'' and ``channel'',
were unusually prominent during this time,
per their time series in Fig.~\ref{fig:storyturb.storyturbulence_twitter_labMT_counts010}.
We are unsure exactly why this artefact appeared for our labMT data set.
We note that in our Twitter $n$-grams project, Storywrangler,
we do not see any anomalous behavior for ``weather'', ``channel'',
or ``weather channel'' in English~\cite{alshaabi2020d,storywrangler2020-08-23a}.
(For Storywrangler, whose development was directly motivated by the findings
of our present paper, we used FastText for language identification of tweets~\cite{alshaabi2020a}.)

Finally, in
Fig.~\ref{fig:storyturb.storyturbulence_twitter_labMT_jsd_daily007},
we show the year-scale JSD time series for our labMT data set
with corrupted words removed.
To be compared with
Fig.~\ref{fig:storyturb.storyturbulence_twitter_labMT_jsd_daily001}C,
we now see a noisy time series more in keeping with the 1-day and 1-week
time scale JSD times series in
Figs.~\ref{fig:storyturb.storyturbulence_twitter_labMT_jsd_daily001}A
and
\ref{fig:storyturb.storyturbulence_twitter_labMT_jsd_daily001}B.
While we cannot be sure that there are no other problems for 
our labMT word list, we have at least been able to
systematically contend with the time series corruptions induced by
changes in Twitter's language detection algorithms.

\section{Concluding remarks}
\label{sec:storyturb.concludingremarks}

We have shown that certain kinds of time series for individual words on Twitter
may be functionally corrupted due
to changes in how Twitter has deployed language detection algorithms over the last decade
coupled with the difficulties of constantly needing to
recognize and adapt to data format changes.
In the absence of the ability to rebuild
these problematic time series from original primary data,
we have demonstrated how 
a systematic, if bespoke, 
method can be developed
to generate a `clean' ensemble of time series.
We repeat that
we do not clean individual time series but rather
remove them entirely from an ensemble.

Anomalies within ensembles of interrelated time series may in general be difficult to discern.
While pursuing other research directions may have uncovered the
same time series problems---our original research interest concerned
lexical turbulence~\cite{pechenick2017a,dodds2020n}---measuring
the divergence between Zipf distributions for days proved powerful here.
Our stumbling upon aberrant time series was helped by
Jensen-Shannon divergence being just one of many divergences that would have worked,
though evidence of
time series problems only arose when we looked beyond short time scales.

We believe our findings should elicit some measure of concern
as they suggest that existing work based on language-specific
time series derived from Twitter may need to be re-examined.
More generally, our work
would support the very reasonable concern any researcher might have about
the long-term integrity of data collected on the fly from social media and other internet
services.
Indeed, our investigations have led us to
rebuild our Twitter database, resulting in important upgrades
for our happiness measurement instrument, Hedonometer,
and the development of our Twitter $n$-gram viewer, Storywrangler.

\acknowledgments
The authors are grateful for the computing resources provided by the
Vermont Advanced Computing Core which was supported in part by NSF
award No. OAC-1827314, and financial support from the Massachusetts
Mutual Life Insurance Company and Google.


\begin{thebibliography}{41}%
\makeatletter
\providecommand \@ifxundefined [1]{%
 \@ifx{#1\undefined}
}%
\providecommand \@ifnum [1]{%
 \ifnum #1\expandafter \@firstoftwo
 \else \expandafter \@secondoftwo
 \fi
}%
\providecommand \@ifx [1]{%
 \ifx #1\expandafter \@firstoftwo
 \else \expandafter \@secondoftwo
 \fi
}%
\providecommand \natexlab [1]{#1}%
\providecommand \enquote  [1]{``#1''}%
\providecommand \bibnamefont  [1]{#1}%
\providecommand \bibfnamefont [1]{#1}%
\providecommand \citenamefont [1]{#1}%
\providecommand \href@noop [0]{\@secondoftwo}%
\providecommand \href [0]{\begingroup \@sanitize@url \@href}%
\providecommand \@href[1]{\@@startlink{#1}\@@href}%
\providecommand \@@href[1]{\endgroup#1\@@endlink}%
\providecommand \@sanitize@url [0]{\catcode `\\12\catcode `\$12\catcode
  `\&12\catcode `\#12\catcode `\^12\catcode `\_12\catcode `\%12\relax}%
\providecommand \@@startlink[1]{}%
\providecommand \@@endlink[0]{}%
\providecommand \url  [0]{\begingroup\@sanitize@url \@url }%
\providecommand \@url [1]{\endgroup\@href {#1}{\urlprefix }}%
\providecommand \urlprefix  [0]{URL }%
\providecommand \Eprint [0]{\href }%
\providecommand \doibase [0]{http://dx.doi.org/}%
\providecommand \selectlanguage [0]{\@gobble}%
\providecommand \bibinfo  [0]{\@secondoftwo}%
\providecommand \bibfield  [0]{\@secondoftwo}%
\providecommand \translation [1]{[#1]}%
\providecommand \BibitemOpen [0]{}%
\providecommand \bibitemStop [0]{}%
\providecommand \bibitemNoStop [0]{.\EOS\space}%
\providecommand \EOS [0]{\spacefactor3000\relax}%
\providecommand \BibitemShut  [1]{\csname bibitem#1\endcsname}%
\let\auto@bib@innerbib\@empty
%</preamble>
\bibitem [{\citenamefont {Osgood}\ \emph {et~al.}(1957)\citenamefont {Osgood},
  \citenamefont {Suci},\ and\ \citenamefont {Tannenbaum}}]{osgood1957a}%
  \BibitemOpen
  \bibfield  {author} {\bibinfo {author} {\bibfnamefont {C.}~\bibnamefont
  {Osgood}}, \bibinfo {author} {\bibfnamefont {G.}~\bibnamefont {Suci}}, \ and\
  \bibinfo {author} {\bibfnamefont {P.}~\bibnamefont {Tannenbaum}},\
  }\href@noop {} {\emph {\bibinfo {title} {The Measurement of Meaning}}}\
  (\bibinfo  {publisher} {University of Illinois},\ \bibinfo {address} {Urbana,
  IL},\ \bibinfo {year} {1957})\BibitemShut {NoStop}%
\bibitem [{\citenamefont {Thurstone}(1959)}]{thurstone1959a}%
  \BibitemOpen
  \bibfield  {author} {\bibinfo {author} {\bibfnamefont {Louis~L.}\
  \bibnamefont {Thurstone}},\ }\href@noop {} {\emph {\bibinfo {title} {The
  Measurement of Values}}}\ (\bibinfo  {publisher} {Univer. Chicago Press},\
  \bibinfo {year} {1959})\BibitemShut {NoStop}%
\bibitem [{\citenamefont {Gooday}(1990)}]{gooday1990a}%
  \BibitemOpen
  \bibfield  {author} {\bibinfo {author} {\bibfnamefont {Graeme}\ \bibnamefont
  {Gooday}},\ }\bibfield  {title} {\enquote {\bibinfo {title} {Precision
  measurement and the genesis of physics teaching laboratories in {V}ictorian
  {B}ritain},}\ }\href@noop {} {\bibfield  {journal} {\bibinfo  {journal} {The
  British journal for the history of science}\ }\textbf {\bibinfo {volume}
  {23}},\ \bibinfo {pages} {25--51} (\bibinfo {year} {1990})}\BibitemShut
  {NoStop}%
\bibitem [{\citenamefont {Gould}(1996)}]{gould1996a}%
  \BibitemOpen
  \bibfield  {author} {\bibinfo {author} {\bibfnamefont {Stephen~Jay}\
  \bibnamefont {Gould}},\ }\href@noop {} {\emph {\bibinfo {title} {The
  Mismeasure of Man}}}\ (\bibinfo  {publisher} {WW Norton \& company},\
  \bibinfo {year} {1996})\BibitemShut {NoStop}%
\bibitem [{\citenamefont {Mohr}\ \emph {et~al.}(2008)\citenamefont {Mohr},
  \citenamefont {Taylor},\ and\ \citenamefont {Newell}}]{mohr2008a}%
  \BibitemOpen
  \bibfield  {author} {\bibinfo {author} {\bibfnamefont {Peter~J.}\
  \bibnamefont {Mohr}}, \bibinfo {author} {\bibfnamefont {Barry~N.}\
  \bibnamefont {Taylor}}, \ and\ \bibinfo {author} {\bibfnamefont {David~B.}\
  \bibnamefont {Newell}},\ }\bibfield  {title} {\enquote {\bibinfo {title}
  {{C}{O}{D}{A}{T}{A} recommended values of the fundamental physical constants:
  2006},}\ }\href@noop {} {\bibfield  {journal} {\bibinfo  {journal} {Journal
  of Physical and Chemical Reference Data}\ }\textbf {\bibinfo {volume} {80}},\
  \bibinfo {pages} {633--1284} (\bibinfo {year} {2008})}\BibitemShut {NoStop}%
\bibitem [{\citenamefont {Golinski}(2008)}]{golinski2008a}%
  \BibitemOpen
  \bibfield  {author} {\bibinfo {author} {\bibfnamefont {Jan}\ \bibnamefont
  {Golinski}},\ }\href@noop {} {\emph {\bibinfo {title} {Making Natural
  Knowledge: Constructivism and the History of Science, with a new preface}}}\
  (\bibinfo  {publisher} {University of Chicago Press},\ \bibinfo {year}
  {2008})\BibitemShut {NoStop}%
\bibitem [{\citenamefont {Pfeffer}\ \emph {et~al.}(2018)\citenamefont
  {Pfeffer}, \citenamefont {Mayer},\ and\ \citenamefont
  {Morstatter}}]{pfeffer2018a}%
  \BibitemOpen
  \bibfield  {author} {\bibinfo {author} {\bibfnamefont {J{\"u}rgen}\
  \bibnamefont {Pfeffer}}, \bibinfo {author} {\bibfnamefont {Katja}\
  \bibnamefont {Mayer}}, \ and\ \bibinfo {author} {\bibfnamefont {Fred}\
  \bibnamefont {Morstatter}},\ }\bibfield  {title} {\enquote {\bibinfo {title}
  {Tampering with {T}witter's sample {A}{P}{I}},}\ }\href@noop {} {\bibfield
  {journal} {\bibinfo  {journal} {EPJ Data Science}\ }\textbf {\bibinfo
  {volume} {7}},\ \bibinfo {pages} {50} (\bibinfo {year} {2018})}\BibitemShut
  {NoStop}%
\bibitem [{\citenamefont {Flake}\ and\ \citenamefont
  {Fried}(2019)}]{flake2019a}%
  \BibitemOpen
  \bibfield  {author} {\bibinfo {author} {\bibfnamefont {Jessica~Kay}\
  \bibnamefont {Flake}}\ and\ \bibinfo {author} {\bibfnamefont {Eiko~I}\
  \bibnamefont {Fried}},\ }\bibfield  {title} {\enquote {\bibinfo {title}
  {Measurement schmeasurement: {Q}uestionable measurement practices and how to
  avoid them},}\ }\href
  {\href{https://psyarxiv.com/hs7wm/}{https://psyarxiv.com/hs7wm/}} {\
  (\bibinfo {year} {2019})}\BibitemShut {NoStop}%
\bibitem [{\citenamefont {Eco}(1989)}]{eco1989a}%
  \BibitemOpen
  \bibfield  {author} {\bibinfo {author} {\bibfnamefont {Umberto}\ \bibnamefont
  {Eco}},\ }\href@noop {} {\emph {\bibinfo {title} {Foucault's Pendulum}}}\
  (\bibinfo  {publisher} {William Weaver, London: Seeker \& Warburg},\ \bibinfo
  {year} {1989})\BibitemShut {NoStop}%
\bibitem [{\citenamefont {Gould}(1999)}]{gould1999a}%
  \BibitemOpen
  \bibfield  {author} {\bibinfo {author} {\bibfnamefont {Stephen~Jay}\
  \bibnamefont {Gould}},\ }\href@noop {} {\emph {\bibinfo {title} {Questioning
  the Millennium: A Rationalist's Guide to a Precisely Arbitrary Countdown
  (Revised Edition)}}}\ (\bibinfo  {publisher} {Crown},\ \bibinfo {year}
  {1999})\BibitemShut {NoStop}%
\bibitem [{\citenamefont {Sobel}(2007)}]{sobel2007a}%
  \BibitemOpen
  \bibfield  {author} {\bibinfo {author} {\bibfnamefont {Dava}\ \bibnamefont
  {Sobel}},\ }\href@noop {} {\emph {\bibinfo {title} {Longitude: The True Story
  of a Lone Genius Who Solved the Greatest Scientific Problem of His Time}}}\
  (\bibinfo  {publisher} {Bloomsbury Publishing, US},\ \bibinfo {year}
  {2007})\BibitemShut {NoStop}%
\bibitem [{\citenamefont {Murthy}(2018)}]{murthy2018a}%
  \BibitemOpen
  \bibfield  {author} {\bibinfo {author} {\bibfnamefont {Dhiraj}\ \bibnamefont
  {Murthy}},\ }\href@noop {} {\emph {\bibinfo {title} {Twitter}}}\ (\bibinfo
  {publisher} {Polity Press Cambridge, UK},\ \bibinfo {year}
  {2018})\BibitemShut {NoStop}%
\bibitem [{\citenamefont {Sakaki}\ \emph {et~al.}(2010)\citenamefont {Sakaki},
  \citenamefont {Okazaki},\ and\ \citenamefont {Matsuo}}]{sakaki2010a}%
  \BibitemOpen
  \bibfield  {author} {\bibinfo {author} {\bibfnamefont {Takeshi}\ \bibnamefont
  {Sakaki}}, \bibinfo {author} {\bibfnamefont {Makoto}\ \bibnamefont
  {Okazaki}}, \ and\ \bibinfo {author} {\bibfnamefont {Yutaka}\ \bibnamefont
  {Matsuo}},\ }\bibfield  {title} {\enquote {\bibinfo {title} {Earthquake
  shakes {T}witter users: {R}eal-time event detection by social sensors},}\
  }in\ \href@noop {} {\emph {\bibinfo {booktitle} {Proceedings of the 19th
  international conference on World wide web}}}\ (\bibinfo {year} {2010})\ pp.\
  \bibinfo {pages} {851--860}\BibitemShut {NoStop}%
\bibitem [{\citenamefont {Lampos}\ and\ \citenamefont
  {Cristianini}(2010)}]{lampos2010tracking}%
  \BibitemOpen
  \bibfield  {author} {\bibinfo {author} {\bibfnamefont {Vasileios}\
  \bibnamefont {Lampos}}\ and\ \bibinfo {author} {\bibfnamefont {Nello}\
  \bibnamefont {Cristianini}},\ }\bibfield  {title} {\enquote {\bibinfo {title}
  {Tracking the flu pandemic by monitoring the social web},}\ }in\ \href@noop
  {} {\emph {\bibinfo {booktitle} {2010 2nd International Workshop on Cognitive
  Information Processing, CIP2010}}}\ (\bibinfo {year} {2010})\ pp.\ \bibinfo
  {pages} {411--416}\BibitemShut {NoStop}%
\bibitem [{\citenamefont {Culotta}(2010)}]{Culotta2010towards}%
  \BibitemOpen
  \bibfield  {author} {\bibinfo {author} {\bibfnamefont {Aron}\ \bibnamefont
  {Culotta}},\ }\bibfield  {title} {\enquote {\bibinfo {title} {Towards
  detecting influenza epidemics by analyzing {T}witter messages},}\ }in\ \href
  {\doibase 10.1145/1964858.1964874} {\emph {\bibinfo {booktitle} {Proceedings
  of the First Workshop on Social Media Analytics}}},\ \bibinfo {series and
  number} {SOMA 10}\ (\bibinfo  {publisher} {Association for Computing
  Machinery},\ \bibinfo {address} {New York, NY, USA},\ \bibinfo {year}
  {2010})\ pp.\ \bibinfo {pages} {115--122}\BibitemShut {NoStop}%
\bibitem [{\citenamefont {Hong}\ and\ \citenamefont
  {Nadler}(2011)}]{hong2011a}%
  \BibitemOpen
  \bibfield  {author} {\bibinfo {author} {\bibfnamefont {Sounman}\ \bibnamefont
  {Hong}}\ and\ \bibinfo {author} {\bibfnamefont {Daniel}\ \bibnamefont
  {Nadler}},\ }\bibfield  {title} {\enquote {\bibinfo {title} {Does the early
  bird move the polls? {T}he use of the social media tool `{T}witter' by {U}{S}
  politicians and its impact on public opinion},}\ }in\ \href@noop {} {\emph
  {\bibinfo {booktitle} {Proceedings of the 12th Annual International Digital
  Government Research Conference: Digital Government Innovation in Challenging
  Times}}}\ (\bibinfo {year} {2011})\ pp.\ \bibinfo {pages}
  {182--186}\BibitemShut {NoStop}%
\bibitem [{\citenamefont {Younus}\ \emph {et~al.}(2011)\citenamefont {Younus},
  \citenamefont {Qureshi}, \citenamefont {Asar}, \citenamefont {Azam},
  \citenamefont {Saeed},\ and\ \citenamefont {Touheed}}]{younus2011average}%
  \BibitemOpen
  \bibfield  {author} {\bibinfo {author} {\bibfnamefont {Arjumand}\
  \bibnamefont {Younus}}, \bibinfo {author} {\bibfnamefont {M~Atif}\
  \bibnamefont {Qureshi}}, \bibinfo {author} {\bibfnamefont {Fiza~Fatima}\
  \bibnamefont {Asar}}, \bibinfo {author} {\bibfnamefont {Muhammad}\
  \bibnamefont {Azam}}, \bibinfo {author} {\bibfnamefont {Muhammad}\
  \bibnamefont {Saeed}}, \ and\ \bibinfo {author} {\bibfnamefont {Nasir}\
  \bibnamefont {Touheed}},\ }\bibfield  {title} {\enquote {\bibinfo {title}
  {What do the average twitterers say: {A} {T}witter model for public opinion
  analysis in the face of major political events},}\ }in\ \href@noop {} {\emph
  {\bibinfo {booktitle} {2011 International Conference on Advances in Social
  Networks Analysis and Mining}}}\ (\bibinfo {organization} {IEEE},\ \bibinfo
  {year} {2011})\ pp.\ \bibinfo {pages} {618--623}\BibitemShut {NoStop}%
\bibitem [{\citenamefont {Bollen}\ \emph {et~al.}(2011)\citenamefont {Bollen},
  \citenamefont {Mao},\ and\ \citenamefont {Zeng}}]{bollen2011b}%
  \BibitemOpen
  \bibfield  {author} {\bibinfo {author} {\bibfnamefont {J.}~\bibnamefont
  {Bollen}}, \bibinfo {author} {\bibfnamefont {H.}~\bibnamefont {Mao}}, \ and\
  \bibinfo {author} {\bibfnamefont {X.-J.}\ \bibnamefont {Zeng}},\ }\bibfield
  {title} {\enquote {\bibinfo {title} {Twitter mood predicts the stock
  market},}\ }\href@noop {} {\bibfield  {journal} {\bibinfo  {journal} {Journal
  of Computational Science}\ }\textbf {\bibinfo {volume} {2}},\ \bibinfo
  {pages} {1--8} (\bibinfo {year} {2011})}\BibitemShut {NoStop}%
\bibitem [{\citenamefont {Pickard}\ \emph {et~al.}(2011)\citenamefont
  {Pickard}, \citenamefont {Pan}, \citenamefont {Rahwan}, \citenamefont
  {Cebrian}, \citenamefont {Crane}, \citenamefont {Madan},\ and\ \citenamefont
  {Pentland}}]{pickard2011time}%
  \BibitemOpen
  \bibfield  {author} {\bibinfo {author} {\bibfnamefont {Galen}\ \bibnamefont
  {Pickard}}, \bibinfo {author} {\bibfnamefont {Wei}\ \bibnamefont {Pan}},
  \bibinfo {author} {\bibfnamefont {Iyad}\ \bibnamefont {Rahwan}}, \bibinfo
  {author} {\bibfnamefont {Manuel}\ \bibnamefont {Cebrian}}, \bibinfo {author}
  {\bibfnamefont {Riley}\ \bibnamefont {Crane}}, \bibinfo {author}
  {\bibfnamefont {Anmol}\ \bibnamefont {Madan}}, \ and\ \bibinfo {author}
  {\bibfnamefont {Alex}\ \bibnamefont {Pentland}},\ }\bibfield  {title}
  {\enquote {\bibinfo {title} {Time-critical social mobilization},}\
  }\href@noop {} {\bibfield  {journal} {\bibinfo  {journal} {Science}\ }\textbf
  {\bibinfo {volume} {334}},\ \bibinfo {pages} {509--512} (\bibinfo {year}
  {2011})}\BibitemShut {NoStop}%
\bibitem [{\citenamefont {Gao}\ \emph {et~al.}(2011)\citenamefont {Gao},
  \citenamefont {Barbier},\ and\ \citenamefont {Goolsby}}]{gao2011harnessing}%
  \BibitemOpen
  \bibfield  {author} {\bibinfo {author} {\bibfnamefont {Huiji}\ \bibnamefont
  {Gao}}, \bibinfo {author} {\bibfnamefont {Geoffrey}\ \bibnamefont {Barbier}},
  \ and\ \bibinfo {author} {\bibfnamefont {Rebecca}\ \bibnamefont {Goolsby}},\
  }\bibfield  {title} {\enquote {\bibinfo {title} {Harnessing the crowdsourcing
  power of social media for disaster relief},}\ }\href@noop {} {\bibfield
  {journal} {\bibinfo  {journal} {IEEE Intelligent Systems}\ }\textbf {\bibinfo
  {volume} {26}},\ \bibinfo {pages} {10--14} (\bibinfo {year}
  {2011})}\BibitemShut {NoStop}%
\bibitem [{\citenamefont {Kursuncu}\ \emph {et~al.}(2019)\citenamefont
  {Kursuncu}, \citenamefont {Gaur}, \citenamefont {Lokala}, \citenamefont
  {Thirunarayan}, \citenamefont {Sheth},\ and\ \citenamefont
  {Arpinar}}]{kursuncu2019a}%
  \BibitemOpen
  \bibfield  {author} {\bibinfo {author} {\bibfnamefont {Ugur}\ \bibnamefont
  {Kursuncu}}, \bibinfo {author} {\bibfnamefont {Manas}\ \bibnamefont {Gaur}},
  \bibinfo {author} {\bibfnamefont {Usha}\ \bibnamefont {Lokala}}, \bibinfo
  {author} {\bibfnamefont {Krishnaprasad}\ \bibnamefont {Thirunarayan}},
  \bibinfo {author} {\bibfnamefont {Amit}\ \bibnamefont {Sheth}}, \ and\
  \bibinfo {author} {\bibfnamefont {I~Budak}\ \bibnamefont {Arpinar}},\
  }\bibfield  {title} {\enquote {\bibinfo {title} {Predictive analysis on
  {T}witter: {T}echniques and applications},}\ }in\ \href@noop {} {\emph
  {\bibinfo {booktitle} {Emerging research challenges and opportunities in
  computational social network analysis and mining}}}\ (\bibinfo  {publisher}
  {Springer},\ \bibinfo {year} {2019})\ pp.\ \bibinfo {pages}
  {67--104}\BibitemShut {NoStop}%
\bibitem [{\citenamefont {Grinberg}\ \emph {et~al.}(2019)\citenamefont
  {Grinberg}, \citenamefont {Joseph}, \citenamefont {Friedland}, \citenamefont
  {Swire-Thompson},\ and\ \citenamefont {Lazer}}]{grinberg2019a}%
  \BibitemOpen
  \bibfield  {author} {\bibinfo {author} {\bibfnamefont {Nir}\ \bibnamefont
  {Grinberg}}, \bibinfo {author} {\bibfnamefont {Kenneth}\ \bibnamefont
  {Joseph}}, \bibinfo {author} {\bibfnamefont {Lisa}\ \bibnamefont
  {Friedland}}, \bibinfo {author} {\bibfnamefont {Briony}\ \bibnamefont
  {Swire-Thompson}}, \ and\ \bibinfo {author} {\bibfnamefont {David}\
  \bibnamefont {Lazer}},\ }\bibfield  {title} {\enquote {\bibinfo {title} {Fake
  news on {T}witter during the 2016 {U}{S} presidential election},}\
  }\href@noop {} {\bibfield  {journal} {\bibinfo  {journal} {Science}\ }\textbf
  {\bibinfo {volume} {363}},\ \bibinfo {pages} {374--378} (\bibinfo {year}
  {2019})}\BibitemShut {NoStop}%
\bibitem [{\citenamefont {Gallagher}\ \emph {et~al.}(2019)\citenamefont
  {Gallagher}, \citenamefont {Stowell}, \citenamefont {Parker},\ and\
  \citenamefont {Foucault~Welles}}]{gallagher2019a}%
  \BibitemOpen
  \bibfield  {author} {\bibinfo {author} {\bibfnamefont {Ryan~J}\ \bibnamefont
  {Gallagher}}, \bibinfo {author} {\bibfnamefont {Elizabeth}\ \bibnamefont
  {Stowell}}, \bibinfo {author} {\bibfnamefont {Andrea~G}\ \bibnamefont
  {Parker}}, \ and\ \bibinfo {author} {\bibfnamefont {Brooke}\ \bibnamefont
  {Foucault~Welles}},\ }\bibfield  {title} {\enquote {\bibinfo {title}
  {Reclaiming stigmatized narratives: {T}he networked disclosure landscape of
  \#{M}e{T}oo},}\ }\href@noop {} {\bibfield  {journal} {\bibinfo  {journal}
  {Proceedings of the ACM on Human-Computer Interaction}\ }\textbf {\bibinfo
  {volume} {3}},\ \bibinfo {pages} {1--30} (\bibinfo {year}
  {2019})}\BibitemShut {NoStop}%
\bibitem [{\citenamefont {Nakov}\ \emph {et~al.}(2019)\citenamefont {Nakov},
  \citenamefont {Ritter}, \citenamefont {Rosenthal}, \citenamefont
  {Sebastiani},\ and\ \citenamefont {Stoyanov}}]{nakov2019a}%
  \BibitemOpen
  \bibfield  {author} {\bibinfo {author} {\bibfnamefont {Preslav}\ \bibnamefont
  {Nakov}}, \bibinfo {author} {\bibfnamefont {Alan}\ \bibnamefont {Ritter}},
  \bibinfo {author} {\bibfnamefont {Sara}\ \bibnamefont {Rosenthal}}, \bibinfo
  {author} {\bibfnamefont {Fabrizio}\ \bibnamefont {Sebastiani}}, \ and\
  \bibinfo {author} {\bibfnamefont {Veselin}\ \bibnamefont {Stoyanov}},\
  }\bibfield  {title} {\enquote {\bibinfo {title} {Sem{E}val-2016 task 4:
  {S}entiment analysis in {T}witter},}\ }\href@noop {} {\bibfield  {journal}
  {\bibinfo  {journal} {arXiv preprint arXiv:1912.01973}\ } (\bibinfo {year}
  {2019})}\BibitemShut {NoStop}%
\bibitem [{\citenamefont {Jackson}\ \emph {et~al.}(2020)\citenamefont
  {Jackson}, \citenamefont {Bailey},\ and\ \citenamefont
  {Foucault~Welles}}]{jackson2020a}%
  \BibitemOpen
  \bibfield  {author} {\bibinfo {author} {\bibfnamefont {Sarah~J.}\
  \bibnamefont {Jackson}}, \bibinfo {author} {\bibfnamefont {Moya}\
  \bibnamefont {Bailey}}, \ and\ \bibinfo {author} {\bibfnamefont {Brooke}\
  \bibnamefont {Foucault~Welles}},\ }\href@noop {} {\emph {\bibinfo {title}
  {\#HashtagActivism: Networks of Race and Gender Justice}}}\ (\bibinfo
  {publisher} {MIT Press},\ \bibinfo {year} {2020})\BibitemShut {NoStop}%
\bibitem [{\citenamefont {Tangherlini}\ \emph {et~al.}(2020)\citenamefont
  {Tangherlini}, \citenamefont {Shahsavari}, \citenamefont {Shahbazi},
  \citenamefont {Ebrahimzadeh},\ and\ \citenamefont
  {Roychowdhury}}]{tangherlini2020a}%
  \BibitemOpen
  \bibfield  {author} {\bibinfo {author} {\bibfnamefont {Timothy~R}\
  \bibnamefont {Tangherlini}}, \bibinfo {author} {\bibfnamefont {Shadi}\
  \bibnamefont {Shahsavari}}, \bibinfo {author} {\bibfnamefont {Behnam}\
  \bibnamefont {Shahbazi}}, \bibinfo {author} {\bibfnamefont {Ehsan}\
  \bibnamefont {Ebrahimzadeh}}, \ and\ \bibinfo {author} {\bibfnamefont
  {Vwani}\ \bibnamefont {Roychowdhury}},\ }\bibfield  {title} {\enquote
  {\bibinfo {title} {An automated pipeline for the discovery of conspiracy and
  conspiracy theory narrative frameworks: {B}ridgegate, {P}izzagate and
  storytelling on the web},}\ }\href@noop {} {\bibfield  {journal} {\bibinfo
  {journal} {PLoS ONE}\ }\textbf {\bibinfo {volume} {15}},\ \bibinfo {pages}
  {e0233879} (\bibinfo {year} {2020})}\BibitemShut {NoStop}%
\bibitem [{\citenamefont {Allen}\ \emph {et~al.}(2020)\citenamefont {Allen},
  \citenamefont {Howland}, \citenamefont {Mobius}, \citenamefont {Rothschild},\
  and\ \citenamefont {Watts}}]{allen2020a}%
  \BibitemOpen
  \bibfield  {author} {\bibinfo {author} {\bibfnamefont {Jennifer}\
  \bibnamefont {Allen}}, \bibinfo {author} {\bibfnamefont {Baird}\ \bibnamefont
  {Howland}}, \bibinfo {author} {\bibfnamefont {Markus}\ \bibnamefont
  {Mobius}}, \bibinfo {author} {\bibfnamefont {David}\ \bibnamefont
  {Rothschild}}, \ and\ \bibinfo {author} {\bibfnamefont {Duncan~J}\
  \bibnamefont {Watts}},\ }\bibfield  {title} {\enquote {\bibinfo {title}
  {Evaluating the fake news problem at the scale of the information
  ecosystem},}\ }\href@noop {} {\bibfield  {journal} {\bibinfo  {journal}
  {Science Advances}\ }\textbf {\bibinfo {volume} {6}},\ \bibinfo {pages}
  {eaay3539} (\bibinfo {year} {2020})}\BibitemShut {NoStop}%
\bibitem [{\citenamefont {Steinert-Threlkeld}\ \emph
  {et~al.}(2015)\citenamefont {Steinert-Threlkeld}, \citenamefont {Mocanu},
  \citenamefont {Vespignani},\ and\ \citenamefont
  {Fowler}}]{steinert2015online}%
  \BibitemOpen
  \bibfield  {author} {\bibinfo {author} {\bibfnamefont {Zachary~C.}\
  \bibnamefont {Steinert-Threlkeld}}, \bibinfo {author} {\bibfnamefont {Delia}\
  \bibnamefont {Mocanu}}, \bibinfo {author} {\bibfnamefont {Alessandro}\
  \bibnamefont {Vespignani}}, \ and\ \bibinfo {author} {\bibfnamefont {James}\
  \bibnamefont {Fowler}},\ }\bibfield  {title} {\enquote {\bibinfo {title}
  {Online social networks and offline protest},}\ }\href@noop {} {\bibfield
  {journal} {\bibinfo  {journal} {EPJ Data Science}\ }\textbf {\bibinfo
  {volume} {4}},\ \bibinfo {pages} {19} (\bibinfo {year} {2015})}\BibitemShut
  {NoStop}%
\bibitem [{\citenamefont {Dodds}\ \emph {et~al.}(2011)\citenamefont {Dodds},
  \citenamefont {Harris}, \citenamefont {Kloumann}, \citenamefont {Bliss},\
  and\ \citenamefont {Danforth}}]{dodds2011e}%
  \BibitemOpen
  \bibfield  {author} {\bibinfo {author} {\bibfnamefont {P.~S.}\ \bibnamefont
  {Dodds}}, \bibinfo {author} {\bibfnamefont {K.~D.}\ \bibnamefont {Harris}},
  \bibinfo {author} {\bibfnamefont {I.~M.}\ \bibnamefont {Kloumann}}, \bibinfo
  {author} {\bibfnamefont {C.~A.}\ \bibnamefont {Bliss}}, \ and\ \bibinfo
  {author} {\bibfnamefont {C.~M.}\ \bibnamefont {Danforth}},\ }\bibfield
  {title} {\enquote {\bibinfo {title} {Temporal patterns of happiness and
  information in a global social network: {H}edonometrics and {T}witter},}\
  }\href@noop {} {\bibfield  {journal} {\bibinfo  {journal} {PLoS ONE}\
  }\textbf {\bibinfo {volume} {6}},\ \bibinfo {pages} {e26752} (\bibinfo {year}
  {2011})}\BibitemShut {NoStop}%
\bibitem [{\citenamefont {Alshaabi}\ \emph
  {et~al.}(2020{\natexlab{a}})\citenamefont {Alshaabi}, \citenamefont
  {Dewhurst}, \citenamefont {Minot}, \citenamefont {Arnold}, \citenamefont
  {Adams}, \citenamefont {Danforth},\ and\ \citenamefont
  {Dodds}}]{alshaabi2020a}%
  \BibitemOpen
  \bibfield  {author} {\bibinfo {author} {\bibfnamefont {Thayer}\ \bibnamefont
  {Alshaabi}}, \bibinfo {author} {\bibfnamefont {David~R.}\ \bibnamefont
  {Dewhurst}}, \bibinfo {author} {\bibfnamefont {Joshua~R.}\ \bibnamefont
  {Minot}}, \bibinfo {author} {\bibfnamefont {Michael~V.}\ \bibnamefont
  {Arnold}}, \bibinfo {author} {\bibfnamefont {Jane~L.}\ \bibnamefont {Adams}},
  \bibinfo {author} {\bibfnamefont {Christopher~M.}\ \bibnamefont {Danforth}},
  \ and\ \bibinfo {author} {\bibfnamefont {Peter~Sheridan}\ \bibnamefont
  {Dodds}},\ }\href@noop {} {\enquote {\bibinfo {title} {The growing
  amplification of social media: Measuring temporal and social contagion
  dynamics for over 150 languages on twitter for 2009--2020},}\ } (\bibinfo
  {year} {2020}{\natexlab{a}}),\ \bibinfo {note} {available online at
  \href{http://arxiv.org/abs/2003.03667}{http://arxiv.org/abs/2003.03667}}\BibitemShut
  {NoStop}%
\bibitem [{\citenamefont {Alshaabi}\ \emph {et~al.}()\citenamefont {Alshaabi},
  \citenamefont {Adams}, \citenamefont {Arnold}, \citenamefont {Minot},
  \citenamefont {Dewhurst}, \citenamefont {Reagan}, \citenamefont {Danforth},\
  and\ \citenamefont {Dodds}}]{alshaabi2020d}%
  \BibitemOpen
  \bibfield  {author} {\bibinfo {author} {\bibfnamefont {Thayer}\ \bibnamefont
  {Alshaabi}}, \bibinfo {author} {\bibfnamefont {Jane~L.}\ \bibnamefont
  {Adams}}, \bibinfo {author} {\bibfnamefont {Michael~V.}\ \bibnamefont
  {Arnold}}, \bibinfo {author} {\bibfnamefont {Joshua~R.}\ \bibnamefont
  {Minot}}, \bibinfo {author} {\bibfnamefont {David~R.}\ \bibnamefont
  {Dewhurst}}, \bibinfo {author} {\bibfnamefont {Andrew~J.}\ \bibnamefont
  {Reagan}}, \bibinfo {author} {\bibfnamefont {Christopher~M.}\ \bibnamefont
  {Danforth}}, \ and\ \bibinfo {author} {\bibfnamefont {Peter~Sheridan}\
  \bibnamefont {Dodds}},\ }\href@noop {} {\enquote {\bibinfo {title}
  {Storywrangler: {A} massive exploratorium for sociolinguistic, cultural,
  socioeconomic, and political timelines using {T}witter},}\ }\BibitemShut
  {NoStop}%
\bibitem [{sto()}]{storywrangler2020-08-23a}%
  \BibitemOpen
  \href@noop {} {}\bibinfo {note}
  {\href{storywrangling.org}{storywrangling.org}, accessed August 23,
  2020}\BibitemShut {NoStop}%
\bibitem [{\citenamefont {Dodds}\ \emph {et~al.}(2015)\citenamefont {Dodds},
  \citenamefont {Clark}, \citenamefont {Desu}, \citenamefont {Frank},
  \citenamefont {Reagan}, \citenamefont {Williams}, \citenamefont {Mitchell},
  \citenamefont {Harris}, \citenamefont {Kloumann}, \citenamefont {Bagrow},
  \citenamefont {Megerdoomian}, \citenamefont {McMahon}, \citenamefont
  {Tivnan},\ and\ \citenamefont {Danforth}}]{dodds2015a}%
  \BibitemOpen
  \bibfield  {author} {\bibinfo {author} {\bibfnamefont {P.~S.}\ \bibnamefont
  {Dodds}}, \bibinfo {author} {\bibfnamefont {E.~M.}\ \bibnamefont {Clark}},
  \bibinfo {author} {\bibfnamefont {S.}~\bibnamefont {Desu}}, \bibinfo {author}
  {\bibfnamefont {M.~R.}\ \bibnamefont {Frank}}, \bibinfo {author}
  {\bibfnamefont {A.~J.}\ \bibnamefont {Reagan}}, \bibinfo {author}
  {\bibfnamefont {J.~R.}\ \bibnamefont {Williams}}, \bibinfo {author}
  {\bibfnamefont {L.}~\bibnamefont {Mitchell}}, \bibinfo {author}
  {\bibfnamefont {K.~D.}\ \bibnamefont {Harris}}, \bibinfo {author}
  {\bibfnamefont {I.~M.}\ \bibnamefont {Kloumann}}, \bibinfo {author}
  {\bibfnamefont {J.~P.}\ \bibnamefont {Bagrow}}, \bibinfo {author}
  {\bibfnamefont {K.}~\bibnamefont {Megerdoomian}}, \bibinfo {author}
  {\bibfnamefont {M.~T.}\ \bibnamefont {McMahon}}, \bibinfo {author}
  {\bibfnamefont {B.~F.}\ \bibnamefont {Tivnan}}, \ and\ \bibinfo {author}
  {\bibfnamefont {C.~M.}\ \bibnamefont {Danforth}},\ }\bibfield  {title}
  {\enquote {\bibinfo {title} {Human language reveals a universal positivity
  bias},}\ }\href@noop {} {\bibfield  {journal} {\bibinfo  {journal} {Proc.
  Natl. Acad. Sci.}\ }\textbf {\bibinfo {volume} {112}},\ \bibinfo {pages}
  {2389--2394} (\bibinfo {year} {2015})},\ \bibinfo {note} {available online at
  \url{http://www.pnas.org/content/112/8/2389}}\BibitemShut {NoStop}%
\bibitem [{hed()}]{hedonometer2020-08-23a}%
  \BibitemOpen
  \href@noop {} {}\bibinfo {note} {\href{hedonometer.org}{hedonometer.org},
  accessed August 23, 2020}\BibitemShut {NoStop}%
\bibitem [{\citenamefont {Alshaabi}\ \emph
  {et~al.}(2020{\natexlab{b}})\citenamefont {Alshaabi}, \citenamefont {Arnold},
  \citenamefont {Minot}, \citenamefont {Adams}, \citenamefont {Dewhurst},
  \citenamefont {Reagan}, \citenamefont {Muhamad}, \citenamefont {Danforth},\
  and\ \citenamefont {Dodds}}]{alshaabi2020b}%
  \BibitemOpen
  \bibfield  {author} {\bibinfo {author} {\bibfnamefont {Thayer}\ \bibnamefont
  {Alshaabi}}, \bibinfo {author} {\bibfnamefont {Michael~V.}\ \bibnamefont
  {Arnold}}, \bibinfo {author} {\bibfnamefont {Joshua~R.}\ \bibnamefont
  {Minot}}, \bibinfo {author} {\bibfnamefont {Jane~L.}\ \bibnamefont {Adams}},
  \bibinfo {author} {\bibfnamefont {David~R.}\ \bibnamefont {Dewhurst}},
  \bibinfo {author} {\bibfnamefont {Andrew~J.}\ \bibnamefont {Reagan}},
  \bibinfo {author} {\bibfnamefont {Roby}\ \bibnamefont {Muhamad}}, \bibinfo
  {author} {\bibfnamefont {Christopher~M.}\ \bibnamefont {Danforth}}, \ and\
  \bibinfo {author} {\bibfnamefont {Peter~Sheridan}\ \bibnamefont {Dodds}},\
  }\href@noop {} {\enquote {\bibinfo {title} {How the world's collective
  attention is being paid to a pandemic: {C}{O}{V}{I}{D}-19 related $n$-gram
  time series for 24 languages on {T}witter},}\ } (\bibinfo {year}
  {2020}{\natexlab{b}}),\ \bibinfo {note} {available online at
  \href{http://arxiv.org/abs/2003.12614}{http://arxiv.org/abs/2003.12614}}\BibitemShut
  {NoStop}%
\bibitem [{\citenamefont {Dewhurst}\ \emph {et~al.}(2020)\citenamefont
  {Dewhurst}, \citenamefont {Alshaabi}, \citenamefont {Arnold}, \citenamefont
  {Minot}, \citenamefont {Danforth},\ and\ \citenamefont
  {Dodds}}]{dewhurst2020a}%
  \BibitemOpen
  \bibfield  {author} {\bibinfo {author} {\bibfnamefont {David~Rushing}\
  \bibnamefont {Dewhurst}}, \bibinfo {author} {\bibfnamefont {Thayer}\
  \bibnamefont {Alshaabi}}, \bibinfo {author} {\bibfnamefont {Michael~V.}\
  \bibnamefont {Arnold}}, \bibinfo {author} {\bibfnamefont {Joshua~R.}\
  \bibnamefont {Minot}}, \bibinfo {author} {\bibfnamefont {Christopher~M.}\
  \bibnamefont {Danforth}}, \ and\ \bibinfo {author} {\bibfnamefont
  {Peter~Sheridan}\ \bibnamefont {Dodds}},\ }\href@noop {} {\enquote {\bibinfo
  {title} {Divergent modes of online collective attention to the
  {C}{O}{V}{I}{D}-19 pandemic are associated with future caseload variance},}\
  } (\bibinfo {year} {2020}),\ \bibinfo {note} {available online at
  \href{http://arxiv.org/abs/2004.03516}{http://arxiv.org/abs/2004.03516}}\BibitemShut
  {NoStop}%
\bibitem [{\citenamefont {Dodds}\ \emph
  {et~al.}(2020{\natexlab{a}})\citenamefont {Dodds}, \citenamefont {Minot},
  \citenamefont {Arnold}, \citenamefont {Alshaabi}, \citenamefont {Adams},
  \citenamefont {Reagan},\ and\ \citenamefont {Danforth}}]{dodds2020n}%
  \BibitemOpen
  \bibfield  {author} {\bibinfo {author} {\bibfnamefont {Peter~Sheridan}\
  \bibnamefont {Dodds}}, \bibinfo {author} {\bibfnamefont {Joshua~R.}\
  \bibnamefont {Minot}}, \bibinfo {author} {\bibfnamefont {Michael~V.}\
  \bibnamefont {Arnold}}, \bibinfo {author} {\bibfnamefont {Thayer}\
  \bibnamefont {Alshaabi}}, \bibinfo {author} {\bibfnamefont {Jane~L.}\
  \bibnamefont {Adams}}, \bibinfo {author} {\bibfnamefont {Andrew~J.}\
  \bibnamefont {Reagan}}, \ and\ \bibinfo {author} {\bibfnamefont
  {Christopher~M.}\ \bibnamefont {Danforth}},\ }\href@noop {} {\enquote
  {\bibinfo {title} {Computational timeline reconstruction of the stories
  surrounding {T}rump: {S}tory turbulence, narrative control, and collective
  chronopathy},}\ } (\bibinfo {year} {2020}{\natexlab{a}}),\ \bibinfo {note}
  {available online at
  \href{https://arxiv.org/abs/2008.07301}{https://arxiv.org/abs/2008.07301}}\BibitemShut
  {NoStop}%
\bibitem [{\citenamefont {Pechenick}\ \emph {et~al.}(2017)\citenamefont
  {Pechenick}, \citenamefont {Danforth},\ and\ \citenamefont
  {Dodds}}]{pechenick2017a}%
  \BibitemOpen
  \bibfield  {author} {\bibinfo {author} {\bibfnamefont {Eitan~A.}\
  \bibnamefont {Pechenick}}, \bibinfo {author} {\bibfnamefont {Chrisopher~M.}\
  \bibnamefont {Danforth}}, \ and\ \bibinfo {author} {\bibfnamefont
  {Peter~Sheridan}\ \bibnamefont {Dodds}},\ }\bibfield  {title} {\enquote
  {\bibinfo {title} {Is language evolution grinding to a halt? {T}he scaling of
  lexical turbulence in {E}nglish fiction suggests it is not},}\ }\href@noop {}
  {\bibfield  {journal} {\bibinfo  {journal} {Journal of Computational
  Science}\ }\textbf {\bibinfo {volume} {21}},\ \bibinfo {pages} {24--37}
  (\bibinfo {year} {2017})},\ \bibinfo {note} {available online at
  \href{http://arxiv.org/abs/1503.03512}{http://arxiv.org/abs/1503.03512}}\BibitemShut
  {NoStop}%
\bibitem [{\citenamefont {Bradley}\ and\ \citenamefont
  {Lang}(1999)}]{bradley1999a}%
  \BibitemOpen
  \bibfield  {author} {\bibinfo {author} {\bibfnamefont {M.~M.}\ \bibnamefont
  {Bradley}}\ and\ \bibinfo {author} {\bibfnamefont {P.~J.}\ \bibnamefont
  {Lang}},\ }\href@noop {} {\emph {\bibinfo {title} {Affective norms for
  English words (ANEW): Stimuli, instruction manual and affective ratings}}},\
  \bibinfo {type} {Technical report C-1}\ (\bibinfo  {institution} {University
  of Florida},\ \bibinfo {address} {Gainesville, FL},\ \bibinfo {year}
  {1999})\BibitemShut {NoStop}%
\bibitem [{\citenamefont {Dodds}\ \emph
  {et~al.}(2020{\natexlab{b}})\citenamefont {Dodds}, \citenamefont {Minot},
  \citenamefont {Arnold}, \citenamefont {Alshaabi}, \citenamefont {Adams},
  \citenamefont {Dewhurst}, \citenamefont {Gray}, \citenamefont {Frank},
  \citenamefont {Reagan},\ and\ \citenamefont {Danforth}}]{dodds2020a}%
  \BibitemOpen
  \bibfield  {author} {\bibinfo {author} {\bibfnamefont {Peter~Sheridan}\
  \bibnamefont {Dodds}}, \bibinfo {author} {\bibfnamefont {Joshua~R.}\
  \bibnamefont {Minot}}, \bibinfo {author} {\bibfnamefont {Michael~V.}\
  \bibnamefont {Arnold}}, \bibinfo {author} {\bibfnamefont {Thayer}\
  \bibnamefont {Alshaabi}}, \bibinfo {author} {\bibfnamefont {Jane~Lydia}\
  \bibnamefont {Adams}}, \bibinfo {author} {\bibfnamefont {David~Rushing}\
  \bibnamefont {Dewhurst}}, \bibinfo {author} {\bibfnamefont {Tyler~J.}\
  \bibnamefont {Gray}}, \bibinfo {author} {\bibfnamefont {Morgan~R.}\
  \bibnamefont {Frank}}, \bibinfo {author} {\bibfnamefont {Andrew~J.}\
  \bibnamefont {Reagan}}, \ and\ \bibinfo {author} {\bibfnamefont
  {Christopher~M.}\ \bibnamefont {Danforth}},\ }\href@noop {} {\enquote
  {\bibinfo {title} {Allotaxonometry and rank-turbulence divergence: {A}
  universal instrument for comparing complex systems},}\ } (\bibinfo {year}
  {2020}{\natexlab{b}}),\ \bibinfo {note} {available online at
  \href{https://arxiv.org/abs/2002.09770}{https://arxiv.org/abs/2002.09770}}\BibitemShut
  {NoStop}%
\bibitem [{\citenamefont {Dodds}\ \emph
  {et~al.}(2020{\natexlab{c}})\citenamefont {Dodds} \emph
  {et~al.}}]{dodds2020g}%
  \BibitemOpen
  \bibfield  {author} {\bibinfo {author} {\bibfnamefont {Peter~Sheridan}\
  \bibnamefont {Dodds}} \emph {et~al.},\ }\href@noop {} {\enquote {\bibinfo
  {title} {Probability-turbulence divergence: {A} tunable allotaxonometric
  instrument for comparing heavy-tailed categorical distributions},}\ }
  (\bibinfo {year} {2020}{\natexlab{c}})\BibitemShut {NoStop}%
\end{thebibliography}
\end{document}